\documentclass[aps,prb,twocolumn,superscriptaddress]{revtex4-1}
\usepackage{amssymb}
\usepackage{graphicx}
\usepackage{amsmath}
\usepackage{bm}
\usepackage{natbib}
\usepackage{xcolor}

\usepackage{hyperref}
\renewcommand{\Re}{\mathrm{Re}}
\renewcommand{\Im}{\mathrm{Im}}
\setcitestyle{square,numbers,sort&compress}

\begin{document}
\title{Deep Learning Topological Invariants of Band Insulators}
\author{Ning Sun}
\thanks{They contribute equally to this work. }
\affiliation{Institute for Advanced Study, Tsinghua University, Beijing, 100084, China}
\author{Jinmin Yi}
\thanks{They contribute equally to this work. }
\affiliation{Institute for Advanced Study, Tsinghua University, Beijing, 100084, China}
\affiliation{Department of Physics, Peking University, Beijing, 100871, China }
\author{Pengfei Zhang}
\affiliation{Institute for Advanced Study, Tsinghua University, Beijing, 100084, China}
\author{Huitao Shen}
\affiliation{Department of Physics, Massachusetts Institute of Technology, Cambridge, Massachusetts 02139, USA}
\author{Hui Zhai}
\affiliation{Institute for Advanced Study, Tsinghua University, Beijing, 100084, China}
\affiliation{Collaborative Innovation Center of Quantum Matter, Beijing, 100084, China}

\date{\today}

\begin{abstract}

In this work we design and train deep neural networks to predict topological invariants for one-dimensional four-band insulators in AIII class whose topological invariant is the winding number, and two-dimensional two-band insulators in A class whose topological invariant is the Chern number. Given Hamiltonians in the momentum space as the input, neural networks can predict topological invariants for both classes with accuracy close to or higher than 90\%, even for Hamiltonians whose invariants are beyond the training data set. Despite the complexity of the neural network, we find that the output of certain intermediate hidden layers resembles either the winding angle for models in AIII class or the solid angle (Berry curvature) for models in A class, indicating that neural networks essentially capture the mathematical formula of topological invariants. Our work demonstrates the ability of neural networks to predict topological invariants for complicated models with local Hamiltonians as the only input, and offers an example that even a deep neural network is understandable.

\end{abstract}

\maketitle

\section{Introduction}

Machine learning has achieved huge success recently in industrial applications. In particular, deep
learning prevails for its performance in several different fields including image recognition and speech transcription \cite{lecun2015deep,krizhevsky2012imagenet,farabet2013learning,tompson2014joint,szegedy2015going,mikolov2011strategies,hinton2012deep,sainath2013deep}. In terms of applications in assisting academic research, aside from analyzing experimental data in high-energy physics \cite{whiteson2009machine,baldi2014searching} and astrophysics \cite{dieleman2015rotation,al2015efficient,ball2010data,ravanbakhsh2017enabling}, progresses have also been made on recognizing phases of matter \cite{wetzel2017unsupervised,nomura2017restricted,rao2017identifying,mano2017phase,ch2018unsupervised,wang2017machine,costa2017principal,hu2017discovering,iakovlev2018supervised,li2017applications,arai2018deep,broecker2017quantum,suchsland2018parameter,venderley2017machine,yoshioka2018machine,zhang2018machine,morningstar2017deep,carrasquilla2017machine,broecker2017machine,ch2017machine,zhang2017quantum,zhang2017machine,ohtsuki2016deep,ohtsuki2017deep,schindler2017probing,ponte2017kernel,wang2016discovering,tanaka2017detection,van2017learning,liu2017self,newphase}, accelerating Monte Carlo simulations \cite{liu2017selfmc,huang2017accelerated,huang2017recommender,xu2017self,liu2017selfmcupdate,shen2018self,nagai2017self}, and extracting relations between many-body wavefunctions, entanglement and neural networks \cite{arsenault2015machine,arsenault2014machine,levine2017deep,deng2017quantum,you2018machine,lu2017separability}. Among these progresses, one challenging and interesting problem is to extract global topological features from local inputs, for instance, by supervised training a neural network, and to understand how the neural network works.

In Ref.~\cite{zhang2018machine}, a convolutional neural network is trained to predict the topological invariant for band insulators with high accuracy. The highlights of that work are two-fold. First, only local Hamiltonians are used as the input and no human knowledge is used as a prior. Second, by analyzing the neural network after training, it is found the formula fitted by the neural network is precisely the same as the mathematical formula for the winding number. However, the limitations of Ref.~\cite{zhang2018machine} are also two-fold. Only one-dimensional models in AIII class whose topological invariants are the winding numbers are considered. Moreover, only two-band models are considered.

In this work, we extend the realm of the previous work to more sophisticated scenarios, including (i) one-dimensional models in AIII class with more than two-bands and (ii) two-dimensional two-band models in A class. We find that in both cases, the neural network can predict topological invariants with high accuracy, even for testing Hamiltonians whose topological numbers are beyond those in the training set. Similar to Ref.~\cite{zhang2018machine}, we use local Hamiltonians as the input and do not feature engineer the input data with any human knowledge. Also, the design of the neural network architecture follows general principles, without specifically making use of the prior understanding of topological invariants. The only knowledge we explicitly exploit about these models is the translational symmetry, as we choose convolutional layers as the building blocks of our neural networks. Convolutional layers respect the translational symmetry by construction and reduce the redundancy in the parameterization \cite{footnote1}.

\begin{figure*}[t]
\centering
\includegraphics[width=\textwidth]{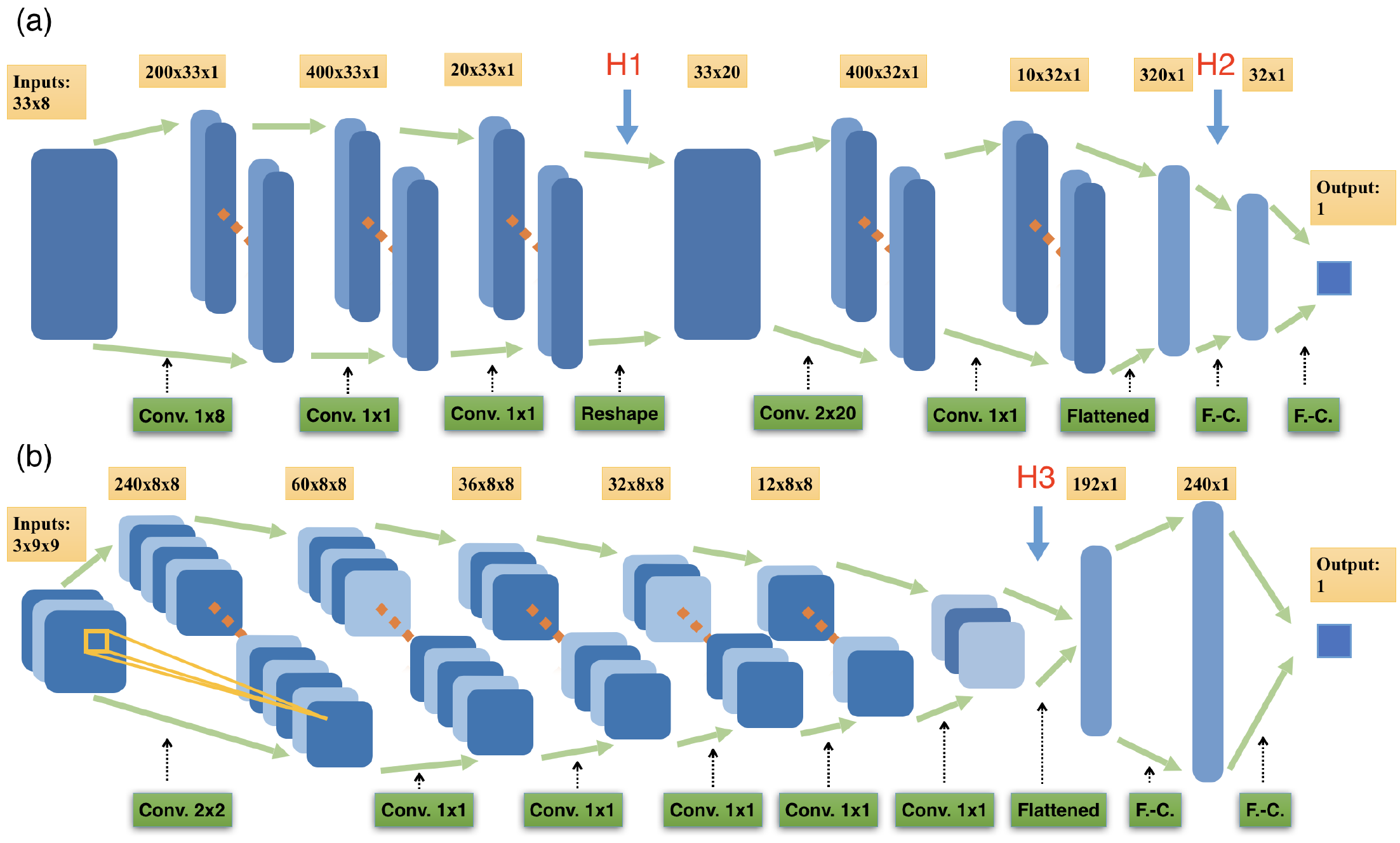}
\caption{The architecture of neural networks used for learning (a) the winding number of one-dimensional AIII class four-band Hamiltonians, and for (b) the Chern number of two-dimensional A class two-band Hamiltonians. In both figures, each linear transformation layer is followed by a subsequent nonlinear ReLU function. The Conv. and F.-C. in the figure denote the convolutinoal layer and the fully-connected layer respectively. The label $a\times b(\times c)$ specifies the dimension of the fully-connected (convolutional) layer. H1, H2 and H3 label layers that we will analyze later.}
\label{architecture}
\end{figure*}

Learning topological invariants of these two models is significantly harder than that in Ref.~\cite{zhang2018machine}, as the mathematical formula of topological invariants in these models are intrinsically more complicated (see Eq.~\eqref{wd1} and Eq.~\eqref{eq:c1}) and the sizes of the input data are much larger. Consequently, to guarantee a good performance, neural networks used in this work are much deeper than the one used in Ref.~\cite{zhang2018machine}. As shown in Fig.~\ref{architecture}, there are more than nine hidden layers in each neural network. Because the neural network becomes more complicated, it becomes more difficult to analyze how the neural network works. Nevertheless, we show that the intermediate output of certain hidden layer is, for case (i) the local winding angle, and for case (ii) the local Berry curvature --- both are the integrands in the mathematical formula of the corresponding topological invariant. In this way, we demonstrate that the complicated function fitted by the neural network is essentially the same as the mathematical formula for the topological invariant.

The paper is organized as follows. In Section II we train a neural network to learn the winding number of one-dimensional four-band models in AIII class. After introducing the model Hamiltonian and the mathematical formula of the winding number, we present our neural network in detail and report its performance. We then analyze the mechanism of why the neural network works. We follow this routine in Section III and show the result for two-dimensional two-band models in A class.

\section{Winding Number with Multiple Bands}

\subsection{Model \label{winding_math}}

Consider a $2d$-band model in one dimension and introduce $ \hat{\Psi}^\dag_k=(\hat{c}^\dag_{1,k},\hat{c}^\dag_{2,k},\dots,\hat{c}^\dag_{2d,k}) $, where $\hat{c}^\dag_{ik}$ is the creation operator for a fermion on $i$-orbital with momentum $k$.
A general one-dimensional four-band Hamiltonian in AIII class can be written as $ \hat{H}=\sum\limits_{k}\hat{\Psi}^\dag_kH(k)\hat{\Psi}_{k} $, where
\begin{align}\label{eq:D(k)}
	H(k)=\begin{pmatrix} 0 & D(k) \\ D^{\dagger}(k) & 0 \end{pmatrix}.
\end{align}
Without loss of generality, here $D(k)\in U(d)$ is a $d$-dimensional unitary matrix \cite{footnote2} and $k\in[-\pi,\pi]$. The topological classification of band Hamiltonians in AIII class is the group $\mathbb{Z}$ \cite{chiu2016classification}. When the model is half-filled, the topological invariant is computed by
\begin{align}
	w=\frac{1}{2\pi}\int_{-\pi}^{\pi}dk\mathrm{Tr}[D^{-1}(k)i\partial_kD(k)]. \label{wd1}
\end{align}
Since $D(k)$ is unitary, it can be diagonalized as $D(k)=V^{\dagger}(k)M(k)V(k)$, where $M(k)$ is a $d$-dimensional diagonal matrix with diagonal elements $\{e^{-i\theta_1(k)}, e^{-i\theta_2(k)},...,e^{-i\theta_d(k)}\}$. Formally, $D(k)$ can also be uniquely decomposed as $D(k)=e^{-i\alpha(k)}\tilde{D}(k)$, where $\tilde{D}(k)\in SU(d)$ is a $d$-dimensional unitary matrix with determinant 1 and $\alpha(k)=\sum_i\theta_i(k)/d\in [-\pi/d,\pi/d)$ is the winding angle at momentum $k$.

To be concrete, we restrict our discussion to $d=2$, which corresponds to four-band models. The winding number formula of Eq.~\eqref{wd1} can then be reduced to
\begin{equation}
	w=\dfrac{1}{\pi}\int_{-\pi}^{\pi}dk\partial_k\alpha(k),
\end{equation}
where $\alpha(k)=(\theta_1(k)+\theta_2(k))/2 \mod \pi$ so that $\alpha(k)\in[-\pi/2,\pi/2)$. The discretized version of the winding number formula is
\begin{align}
	w &=\dfrac{1}{\pi}\sum_{l=1}^{L}\Delta\alpha(k_l) \notag\\
		&=\dfrac{1}{\pi}\sum_{l=1}^{L}[\alpha(k_{l+1})-\alpha(k_l)]\mod \pi, \label{wd_dis}
\end{align}
where $ k_i $, $ i=1,\ldots,L $ are distributed uniformly in the Brillouin zone and $\Delta\alpha(k)\in[-\pi/2,\pi/2)$.

\subsection{Neural Network Performance}

\begin{figure}[t]
	\centering
	\includegraphics[width=1.\columnwidth]{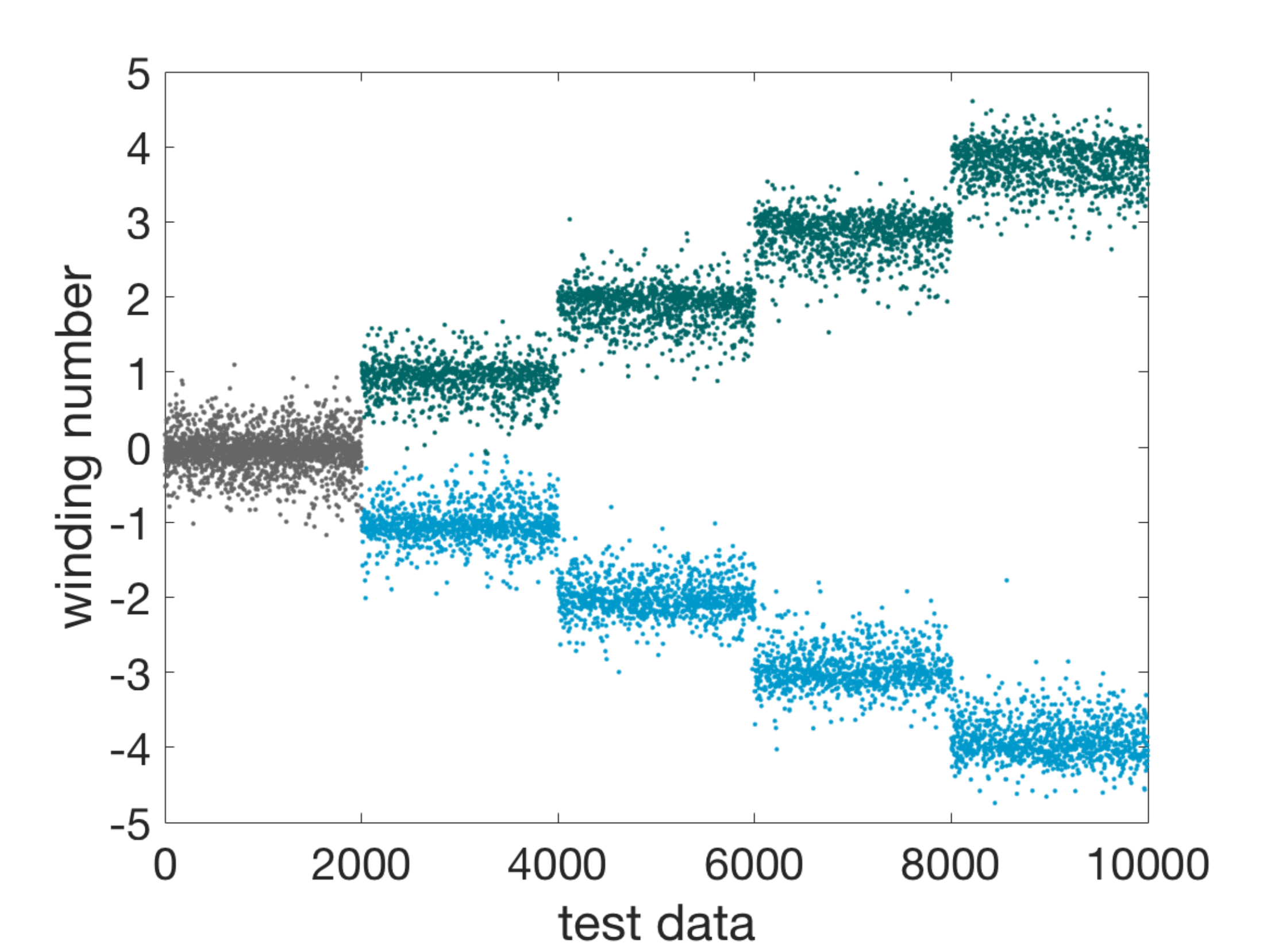}
	\caption{The test data set contains $10^4$ Hamiltonians which are labeled from 1 to 10000. Hamiltonians labeled from $2000i$ to $2000(i+1)$ have winding number $\pm i$, with different colors distinguishing $+i$ from $-i$. The vertical axis shows the winding number (direct output) predicted by the neural network.}\label{fig:winding_test}
\end{figure}

\begin{table}[t]
	\centering
	\caption{The accuracy of the neural network prediction on test Hamiltonians with winding numbers $w=0, \pm1, \pm2, \pm3, \pm4$ respectively. }\label{table:winding-sr}
	\begin{ruledtabular}
	\begin{tabular}{*{8}c}
		& $w$ & 0 & $\pm1$ & $\pm2$ & $\pm3$ & $\pm4$ & \\ \hline
		& Accuracy & 97\% & 96\% & 96\% & 95\% & 93\% &
	\end{tabular}
	\end{ruledtabular}
\end{table}

Since the neural network can only take discrete input, we first discretize the entire Brillouin zone uniformly into $L$ points $\{k_l\in[-\pi,\pi)|l=1,\ldots,L+1\}$ by choosing $k_l=2\pi(l-1)/L$.  At each point, since the Hamiltonian is determined by the $2\times 2$ matrix $D(k)$, we denote its four elements as $D_{11},D_{12},D_{21},D_{22}$. The input data is therefore a $8\times (L+1)$-dimensional matrix of the following form
\begin{align}
	\begin{pmatrix}
		\Re[D_{11}(0)] & \Re[D_{11}(2\pi/L)] & \cdots & \Re[D_{11}(2\pi)] \\
		\Im[D_{11}(0)] & \Im[D_{11}(2\pi/L)] & \cdots & \Im[D_{11}(2\pi)] \\
		\Re[D_{12}(0)] & \Re[D_{12}(2\pi/L)] & \cdots & \Re[D_{12}(2\pi)] \\
		\Im[D_{12}(0)] & \Im[D_{12}(2\pi/L)] & \cdots & \Im[D_{12}(2\pi)] \\
		\Re[D_{21}(0)] & \Re[D_{21}(2\pi/L)] & \cdots & \Re[D_{21}(2\pi)] \\
		\Im[D_{21}(0)] & \Im[D_{21}(2\pi/L)] & \cdots & \Im[D_{21}(2\pi)] \\
		\Re[D_{22}(0)] & \Re[D_{22}(2\pi/L)] & \cdots & \Re[D_{22}(2\pi)] \\
		\Im[D_{22}(0)] & \Im[D_{22}(2\pi/L)] & \cdots & \Im[D_{22}(2\pi)]
	\end{pmatrix}
\end{align}
In the following, we set $L=32$.

The structure of the deep neural network is shown in Fig.~\ref{architecture} (a). It first contains several convolutional layers with kernel sizes marked in the figure, which are followed by two fully-connected layers leading to the final output. In each layer, a linear mapping is followed by a nonlinear ReLU function. We feed the neural network with a set of $3\times 10^4$ discretized training Hamiltonians with winding number $\{0,\pm1,\pm2, \pm 3\}$ for supervised training.

To compute accuracy, the final winding number is taken as the closest integer of the numerical value predicted by the network. It is considered as a correct prediction if the rounded integer matches the value computed by Eq.~\eqref{wd_dis}. The accuracy of this neural network is shown in TABLE \ref{table:winding-sr}. After training, the neural network achieves a prediction accuracy of 96\% on Hamiltonians with winding numbers $\{0,\pm1,\pm2,\pm 3\}$ in a separate test data set, and an accuracy of more than 90\% on Hamiltonians with winding number of $\{\pm4\}$ that are beyond the training set. The numerical values of the winding number predicted for each Hamiltonian in the test set are shown in Fig.~\ref{fig:winding_test}.

\begin{figure}[t]
	\centering
	\includegraphics[width=1.\columnwidth]{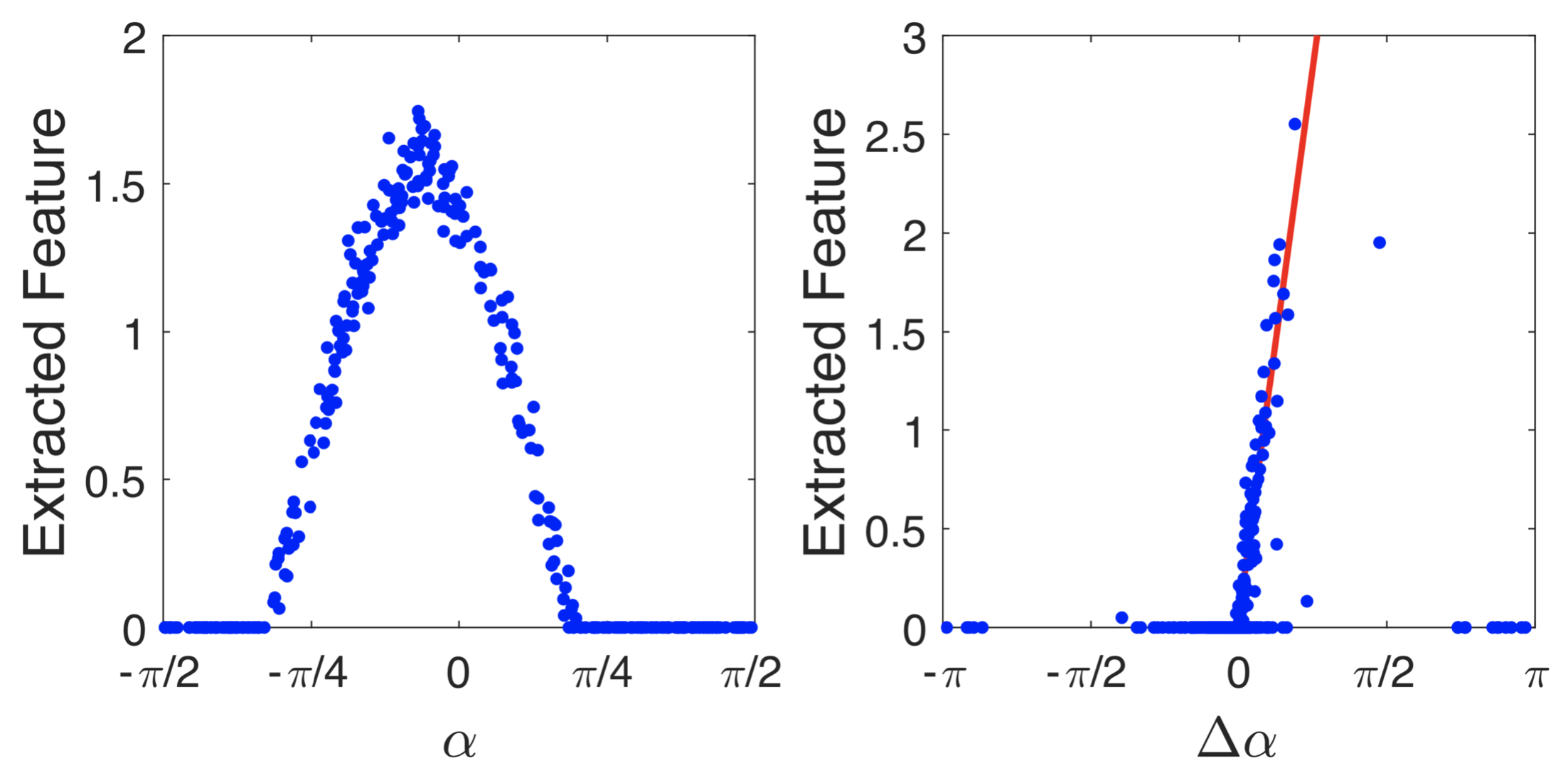}
	\caption{Extracted features of the hidden layers. (a) The intermediate output $r_i$ which is a typical row of the layer marked by H1 in Fig.~\ref{architecture} v.s. the corresponding exact value of $\alpha(k_i)$ for the input Hamiltonian. Other rows exhibit similar behavior which is not shown. (b) The intermediate output $v_i$ which is a typical row of the layer marked by H2 v.s. the corresponding exact value of $\Delta\alpha(k_i)$. Other rows exhibit similar behavior which is not shown. In both figures, the results of 5 different test Hamiltonians are plotted so that there are $5(L+1)$ and $5L$ points in total respectively.  }\label{analysis_winding}
\end{figure}

\subsection{Neural Network Analysis}

To see why the neural network excels predicting the topological winding number, it is illuminating to check whether the complicated function fitted by the neural network is consistent with the mathematical formula Eq.~\eqref{wd_dis} introduced above. We open up the neural network at H1 and H2 marked in Fig.~\ref{architecture} by feeding test Hamiltonians into the neural network and plotting intermediate outputs at H1 and H2 separately. Notice that, the output of H1 is of dimension $(L+1)\times20$, while the dimension of H2 is $L\times10$. Each row of H1 can be interpreted as a vector $r\in\mathbb{R}^{L+1}$, and each row of H2 can be interpreted as vector $v\in\mathbb{R}^{L}$. They respectively have the same dimension as the discretized $\alpha(k)$ and $\Delta\alpha(k)$ defined in Sec.~\ref{winding_math}.
On the other hand, the exact value of $\alpha(k)$ and $\Delta\alpha(k)$ of the corresponding Hamiltonian can also be obtained directly according to the definition in Sec.~\ref{winding_math}.
In Fig.~\ref{analysis_winding}(a) we plot $\{(\alpha(k_i),r_i)|i=1,\ldots,L+1\}$, where $r_i$ is the $i$-th component of a selected row of H1, for various $k_i$ and input Hamiltonians. The plot for H2 in Fig.~\ref{analysis_winding}(b) is similar where $\{(\Delta\alpha(k_i),v_i)|i=1,\ldots,L\}$ are plotted.

As can be seen in Fig.~\ref{analysis_winding}(a), the intermediate output at H1 is approximately piecewise linear with $\alpha$, implying that this row of neuron successfully extracts the winding angle $\alpha$ within some range. Other rows of neurons extracts winding angles at different ranges. In Fig.~\ref{analysis_winding}(b), the intermediate output at H2 is approximately linear with $\Delta\alpha$ within some range, and each row of neuron functions as a $\Delta\alpha$ extractor for different ranges of $\Delta\alpha$. Although their ranges may overlap with each other or have different slopes in their linear relations with the exact $\Delta\alpha$, a linear combination of these extractors with correct coefficients in the following fully-connected layer can easily lead to a function proportional to $\Delta\alpha$ at all ranges. In this way, the winding number is calculated essentially the same way as that using the mathematical formula Eq.~\eqref{wd_dis}.

As emphasized in Sec.~\ref{winding_math}, it is important to notice the input Hamiltonian can be written as the product of a phase factor and a $SU(d)$ matrix. The $SU(d)$ matrix does not play any role in determining the winding number and only the phase factor matters. It is quite impressive that the neural network successfully distills the phase factor from the irrelevant $SU(d)$ part.

\section{Chern Number in Two Dimensions}
\subsection{Model}

Consider a two-band model in two dimensions and introduce $ \hat{\Psi}^\dag_k=(\hat{c}^\dag_{1,{\bf k}},\hat{c}^\dag_{2,{\bf k}}) $,
where $\hat{c}^\dag_{i,{\bf k}}$ is the creation operator for a fermion on $i$-orbital with momentum ${\bf k}=(k_x,k_y)$. A general two-dimensional two-band Hamiltonian in A class can be written as $ \hat{H}=\sum\limits_{{\bf k}}\hat{\Psi}^\dag_{\bf k}H({\bf k})\hat{\Psi}_{{\bf k}} $,
where
\begin{align}
	H(\mathbf{k})=\mathbf{h}(\mathbf{k})\cdot\boldsymbol{\sigma}=h_x(\mathbf{k})\sigma_x+h_y(\mathbf{k})\sigma_y+h_z(\mathbf{k})\sigma_z.
\end{align}
Here $\boldsymbol{\sigma}=(\sigma_x,\sigma_y,\sigma_z)$ is a vector of Pauli matrices.
Without loss of generality, we can take $|\mathbf{h}(\mathbf{k})|=1$ as the normalization \footnote{This is similar to that $D(k)$ is taken as the unitary matrix in the previous case of the winding number, because we can always take flat-band approximation for an insulator without changing its band topology.}. In two dimensions, the Chern number can be computed as
\begin{align}\label{eq:c1}
	C=\dfrac{1}{2\pi}\int_{T^2}d^2\mathbf{k}F_{xy}(\mathbf{k}),
\end{align}
where $T^2$ is the torus of the Brillouin zone and
\begin{equation}\label{eq:c1AF}
A_{\mu}(\mathbf{k}) = i\langle u(\mathbf{k})|\partial_{\mu} u(\mathbf{k})\rangle, \ F_{\mu\nu}(\mathbf{k}) = \partial_{\mu}A_{\nu}-\partial_{\nu}A_{\mu}.
\end{equation}
Here we assume the model is half-filled so that $|u(\mathbf{k})\rangle$ is the energy eigenstate with the lower energy $H({\bf k})|u(\mathbf{k})\rangle=-|u(\mathbf{k})\rangle$.  The integrand in Eq.~\eqref{eq:c1} is then the Berry curvature of the lower band. For discretized lattices, the Berry curvature and the Chern number can be defined through the Wilson-loop approach, as is elaborated in the Appendix.


\begin{figure}[t]
	\centering
	\includegraphics[width=1.\columnwidth]{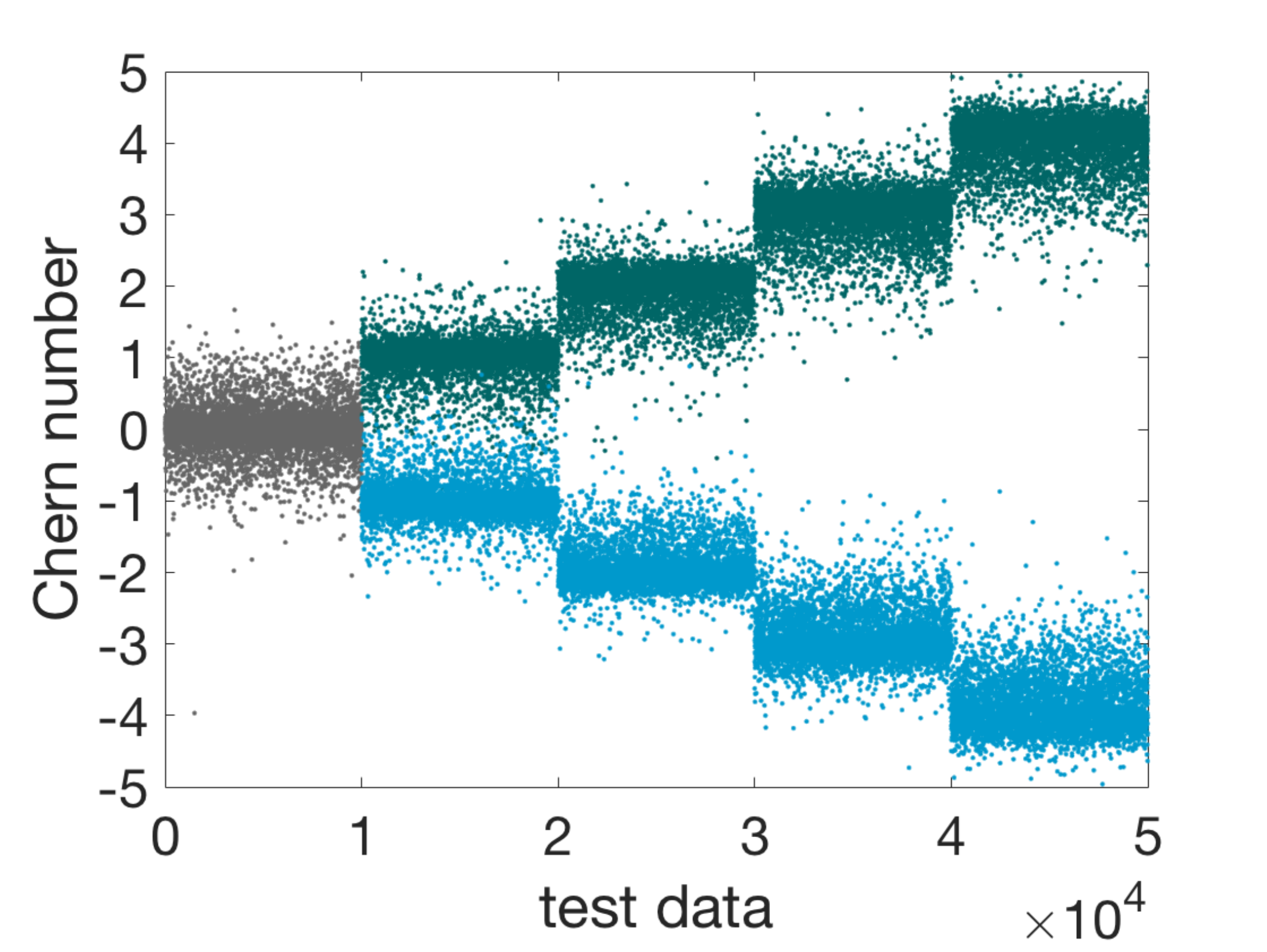}
	\caption{The test data set contains $5\times 10^4$ Hamiltonians which are labeled from $1$ to $5\times 10^4$. The data labeled from $i+1$ to $(i+1)10^4$ has Chern number $ \pm i $, with different colors distinguishing $+i$ from $-i$. The vertical axis shows the Chern number (direct output) predicted by the neural network. }
	\label{fig:Chern_test}
\end{figure}

\begin{table}[t]
	\centering
	\caption{The accuracy of the neural network prediction on test Hamiltonians with Chern numbers $C=0, \pm1, \pm2, \pm3, \pm4$ respectively. }
	\label{table:chern-sr}
	\begin{ruledtabular}
	\begin{tabular}{*{8}c}
		& $C$ & $0$ & $\pm1$ & $\pm2$ & $\pm3$ & $\pm4$ & \\ \hline
		& Accuracy & 93\% & 92\% & 90\% & 86\% & 85\% &
	\end{tabular}
	\end{ruledtabular}
\end{table}

\subsection{Neural Network Performance}

The input data are Hamiltonians in the discretized Brillouin zone, i.e., $3\times (L+1)\times (L+1)$ tensors $\begin{pmatrix}\mathcal{H}_x, & \mathcal{H}_y, & \mathcal{H}_z \end{pmatrix}$ with
\begin{align}
	\mathcal{H}_{\mu} &=
	\begin{pmatrix}
		h_{\mu}(0,0) & h_{\mu}(0,\frac{2\pi}{L}) & \cdots & h_{\mu}(0,2\pi) \\
		h_{\mu}(\frac{2\pi}{L},0) & h_{\mu}(\frac{2\pi}{L},\frac{2\pi}{L}) & \cdots & h_{\mu}(\frac{2\pi}{L},2\pi) \\
		\vdots & \vdots & \ddots & \vdots \\
		h_{\mu}(2\pi,0) & h_{\mu}(2\pi,\frac{2\pi}{L}) & \cdots & h_{\mu}(2\pi,2\pi) \\
	\end{pmatrix}.
\end{align}
The corresponding Chern numbers are calculated using the method presented in the Appendix. In the following, we take $L=8$.

The structure of the neural network is shown in Fig.~\ref{architecture}(b) which is similar to that used for the winding number. We feed the neural network with  $ 10^4$ randomly generated Hamiltonians with Chern numbers limited to $\{0,\pm1,\pm2\}$. The accuracy here is computed similarly to before by rounding the final output of the network to the closet integer. After training, the neural network can achieve an accuracy of $92\%$ on Hamiltonians with Chern numbers $C\in \{0,\pm1,\pm2\}$, an accuracy of $84\%$ on Hamiltonians with Chern numbers $\pm3$ and an accuracy of $85\%$ on Hamiltonians with Chern numbers $\pm4$. These results are shown in Fig.~\ref{fig:Chern_test} and are summarized in TABLE \ref{table:chern-sr}.

\begin{figure}[t]
	\centering
		\includegraphics[width=1.\columnwidth]{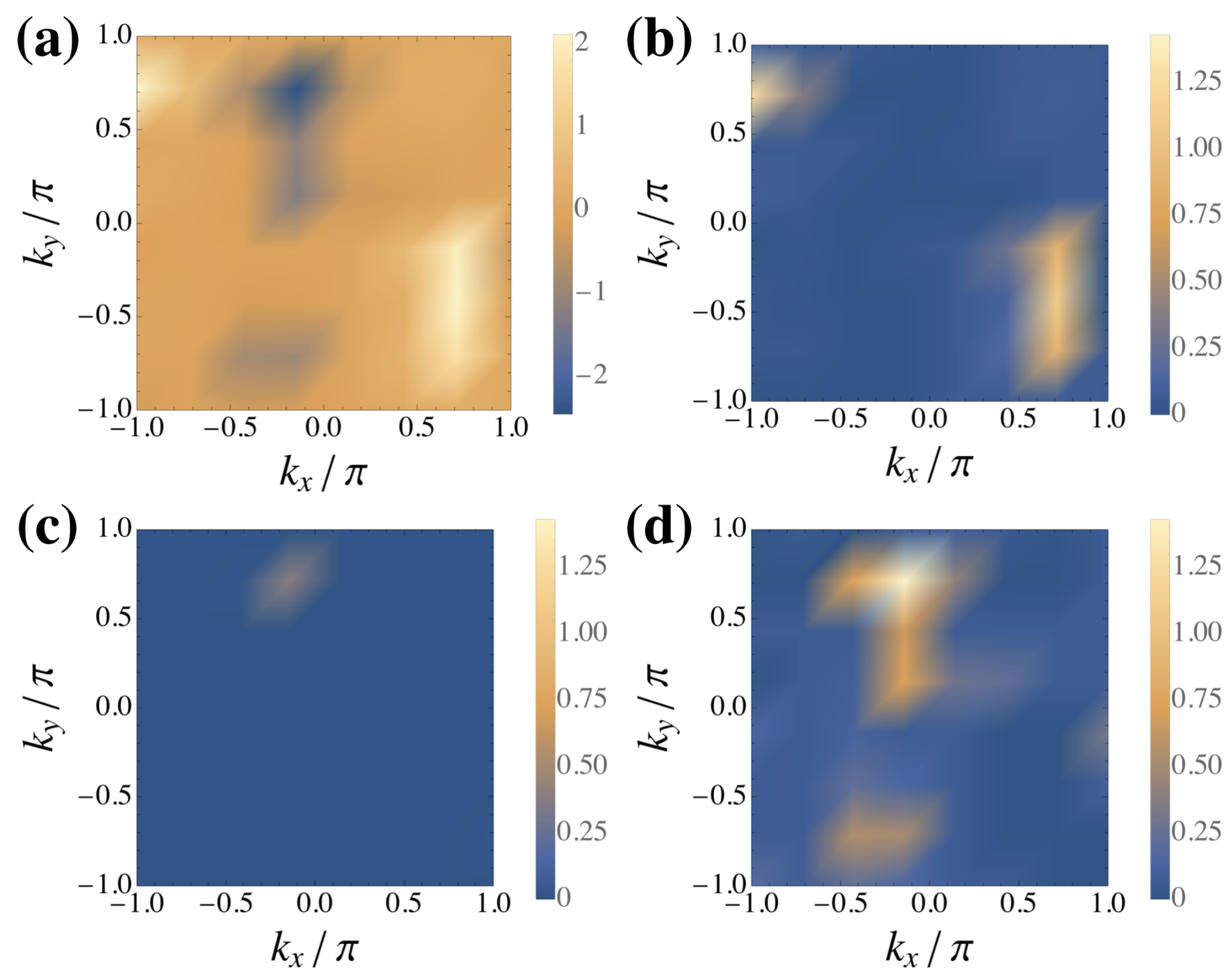}
	\caption{(a) The calculated Berry curvature for a test Hamiltonian in the first Brillouin zone . (b-d) The corresponding intermediate outputs at the layer marked by H3 in Fig. \ref{architecture}(b) before the fully-connected layers. Notice that the output is a 3-tensor, (b), (c) and (d) corresponds to three different components of the 3-tensor. } \label{analysis_Berry}
\end{figure}

\subsection{Neural Network Analysis}

We feed the neural network with a Hamiltonian in the test data set and plot the intermediate output of the last convolutional layer (marked by H3 in Fig.~\ref{architecture}(b)) in Fig.~\ref{analysis_Berry}(b-d). The output consists of three layers of $L\times L$ matrices, which are respectively shown in Fig.~\ref{analysis_Berry}(b), (c) and (d). They should be compared with the exact Berry curvature for the corresponding Hamiltonian shown in Fig.~\ref{analysis_Berry}(a). Since the intermediate output is positive due to nature of the ReLU function while the Berry curvature are generally positive somewhere and negative elsewhere, the intermediate output reproduces the positive part of the Berry curvature in one layer (Fig. \ref{analysis_Berry}(b)) and the negative part in another layer (Fig. \ref{analysis_Berry}(c)). The remaining third layer is almost irresponsive (Fig. \ref{analysis_Berry}(d)). This result shows the neural network compute the topological invariant by first computing local Berry curvatures in the momentum space and then adding them together, which is essentially the same as Eq.~\eqref{eq:c1}.

\section{Summary}

In summary, we have trained deep neural networks to predict the winding number of one-dimensional four-band models in AIII class and the Chern number of two-dimensional two-band models in A class. In addition to the high prediction accuracies after the training, it is understood that deep neural networks essentially fit the mathematical formula for both topological invariants. In the first case, the network successfully distills the $ U(d) $ phase factors of Hamiltonians between two successive momenta and discards the $SU(d)$ degrees of freedom that is redundant in determining the topology. In the second case, the network successfully extracts the Berry curvature in momentum space. Our work provides an explicit example that even a complicated deep neural network can be understood. Our work can be further combined with ab initio calculations, and paves the way to the direct prediction of topological properties of real materials using machine learning.

\appendix

\section{Chern number in discrete spaces}\label{appendix:chern}

\begin{figure}[t]
	\centering
		\includegraphics[width=0.8\columnwidth]{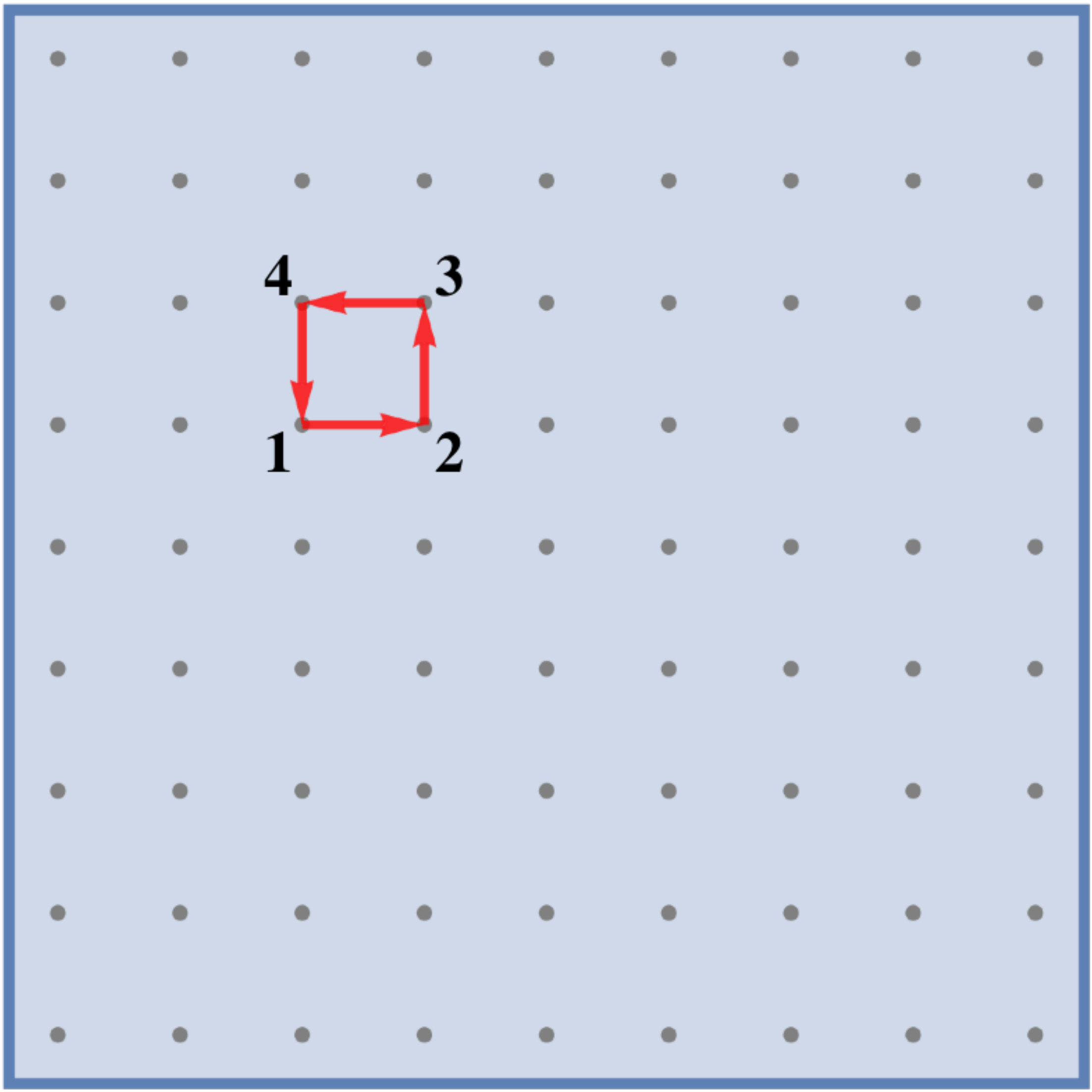}
	\caption{Schematic of discretized two-dimensional parameter space and the Wilson loop. Numbers label the ordering of the loop. }\label{fig:discretized2d}
\end{figure}

The continuous version of Chern number and Berry curvature is defined in Eq.~\eqref{eq:c1AF} in the main text. To introduce the discrete version of Chern number, it is convenient to first define the Berry curvature in discrete spaces\cite{Discretized2005}. The Chern number is then the summation of Berry curvatures in the space.

The definition of the Berry curvature and the Chern number in discrete spaces, and the procedure for computing them are outlined as follows.

1. Discretize a two-dimensional parameter space as $L\times L$ sites. With periodic boundary condition by identifying sites at the boundary, there are $L\times L$ plaquettes in total. In our setting, sites are labeled as $\mathbf{k}=(k_x,k_y)$. For uniform discretization, the area of each plaquette is $ s(\mathbf{k})=\Delta k_{x}\Delta k_{y} $, where $\Delta k_x$ and $\Delta k_y$ is the distance of neighboring sites along $k_x$ and $k_y$ respectively.

2. At each site $\mathbf{k}=(k_x,k_y)$ in the discretized two-dimensional parameter space, diagonalize the Hamiltonian $H(\mathbf{k})=V(\mathbf{k})D(\mathbf{k})V^{\dagger}(\mathbf{k})$ to obtain the eigenstates of the $n$-th band $|u^{(n)}(\mathbf{k})\rangle$. $D(\mathbf{k})$ is a diagonal matrix with its diagonal elements the eigenenergy of each band.

3. All four vertices in each plaquette construct an ordered loop, called the Wilson loop.

	(a). Compute the ordered inner product of the eigenstates along the ordered loop in each plaquette. Specifically, define
	\begin{align*}
		U_{12}=V^{\dagger}(\mathbf{k}_2)V(\mathbf{k}_1),\  U_{23}=V^{\dagger}(\mathbf{k}_3)V(\mathbf{k}_2), \\
		U_{34}=V^{\dagger}(\mathbf{k}_4)V(\mathbf{k}_3),\  U_{41}=V^{\dagger}(\mathbf{k}_1)V(\mathbf{k}_4),
	\end{align*}
	
	(b). Define $ \mathcal{U}_{ij}=\mathrm{diag}(U_{ij}) $, where $\mathrm{diag}(\dots)$ means to extract the diagonal elements and construct a diagonal matrix. That is, $(\mathcal{U}_{ij})_{mn}=\delta_{mn}(U_{ij})_{nn}$.

	(c). Define $ T_{\text{loop}}(\mathbf{k}_1)=\mathcal{U}_{41}\mathcal{U}_{34}\mathcal{U}_{23}\mathcal{U}_{12} $. $-i\log T(\mathbf{k}_1)$ is the (non-Abelian) Berry curvature at the plaqutte labeled $\mathbf{k}_1$. Define $ \theta_n(\mathbf{k})= -i\log [T_{\text{loop}}(k_i,k_j)]_{nn} $ and the Berry curvature of the $n$-th band $ \mathcal{F}_{xy}^{(n)} $
	\begin{equation}
	\mathcal{F}_{xy}^{(n)}(\mathbf{k})=\theta_n(\mathbf{k})/s(\mathbf{k}). \label{eq:af}
	\end{equation}
	
4. The Chern number is the summation of the Berry curvature of all plaquettes. Define $c_n$ as the Chern number of the $n$-th band:
\begin{align}
c_n &= \frac{1}{2\pi}\sum_{i=1}^{L\times L} \theta(\mathbf{k}_i) \notag\\
&=\frac{1}{2\pi}\sum_{i=1}^{L}\sum_{j=1}^{L}-i\log T_{\text{loop}}^{(nn)}(k_i,k_j). \label{eq:ac}
\end{align}
	
It can be verified that the Chern number defined above is quantized and gauge invariant. For a model defined in the continuous space but whose Chern number is computed only on discretized points in the continuous space, Equation \eqref{eq:ac} gives the same result as Eq.~\eqref{eq:c1} if the discretization is dense enough. Hence Eq.~\eqref{eq:af} and \eqref{eq:ac} can be seen as the generalization of the Berry curvature and the Chern number to discrete spaces.

\end{document}